\documentclass{article}

\usepackage[preprint,nonatbib]{neurips_2022}

\usepackage[utf8]{inputenc}
\usepackage[T1]{fontenc}    
\usepackage{hyperref}       
\usepackage{bm}
\usepackage{url}            
\usepackage{booktabs}       
\usepackage{amsmath} 
\usepackage{amsfonts}       
\usepackage{nicefrac}       
\usepackage{microtype}      
\usepackage{xcolor}         
\usepackage{graphicx}
\usepackage{algorithm}
\usepackage{algorithmicx}
\usepackage{algpseudocode}

\usepackage[sorting=none]{biblatex}
\title{Prune and distill: similar reformatting of image information along rat visual cortex and deep neural networks}

\author{%
    Paolo Muratore\\
    International School for Advanced Studies\\
    \texttt{pmurator@sissa.it}\\
    \And
    Sina Tafazoli\\
    Princeton Neuroscience Institute\\
    International School for Advanced Studies\\
    \texttt{tafazoli@princeton.edu}\\
    \And
    Eugenio Piasini\\
    International School for Advanced Studies\\
    \texttt{epiasini@sissa.it}\\
    \And
    Alessandro Laio\\
    International School for Advanced Studies\\
    \texttt{laio@sissa.it}\\
    \And
    Davide Zoccolan\\
    International School for Advanced Studies\\
    \texttt{zoccolan@sissa.it}\\
}

\begin{document}

\maketitle

\begin{abstract}

Visual object recognition has been extensively studied in both neuroscience and computer vision. Recently, the most popular class of artificial systems for this task, deep convolutional neural networks (CNNs), has been shown to provide excellent models for its functional analogue in the brain, the ventral stream in visual cortex. This has prompted questions on what, if any, are the common principles underlying the reformatting of visual information as it flows through a CNN or the ventral stream. Here we consider some prominent statistical patterns that are known to exist in the internal representations of either CNNs or the visual cortex and look for them in the other system. We show that intrinsic dimensionality (ID) of object representations along the rat homologue of the ventral stream presents two distinct expansion-contraction phases, as previously shown for CNNs. Conversely, in CNNs, we show that training results in both distillation and active pruning (mirroring the increase in ID) of low- to middle-level image information in single units, as representations gain the ability to support invariant discrimination, in agreement with previous observations in rat visual cortex. Taken together, our findings suggest that CNNs and visual cortex share a similarly tight relationship between dimensionality expansion/reduction of object representations and reformatting of image information.

\end{abstract}

\section{Introduction}
\label{SEC:Introduction}

Deep Convolutional Neural Networks (CNNs) currently stand as our best class of models of visual processing in the brain \cite{kriegeskorte2015deep, yamins2016using,lindsay2021convolutional}, showing success in: (1) predicting the tuning of individual neurons at various stages of the primate ventral stream \cite{yamins2014performance}; (2) accounting for their ability to encode a variety of object properties \cite{hong2016explicit}; and (3) controlling their activity  via synthetic  stimuli inferred through  model inversion \cite{ponce2019evolving, bashivan2019neural}.  This suggests  that the objective-optimization framework of deep learning offers a parsimonious explanation of the inner workings of complex, hierarchical brain circuits \cite{richards2019deep}, although the latter are likely shaped by very different learning processes (e.g., unsupervised adaptation to the spatiotemporal statistics of the visual input \cite{li2008unsupervised, matteucci2020unsupervised}. Despite this  success, key differences between biological and artificial hierarchical networks exist (e.g., in sensitivity  to noise or adversarial examples \cite{goodfellow2014explaining, kurakin2018adversarial}), possibly highlighting core dissimilarities in how information is processed in the two systems.

\begin{figure}[t]
    \centering
    \includegraphics[width=13cm]{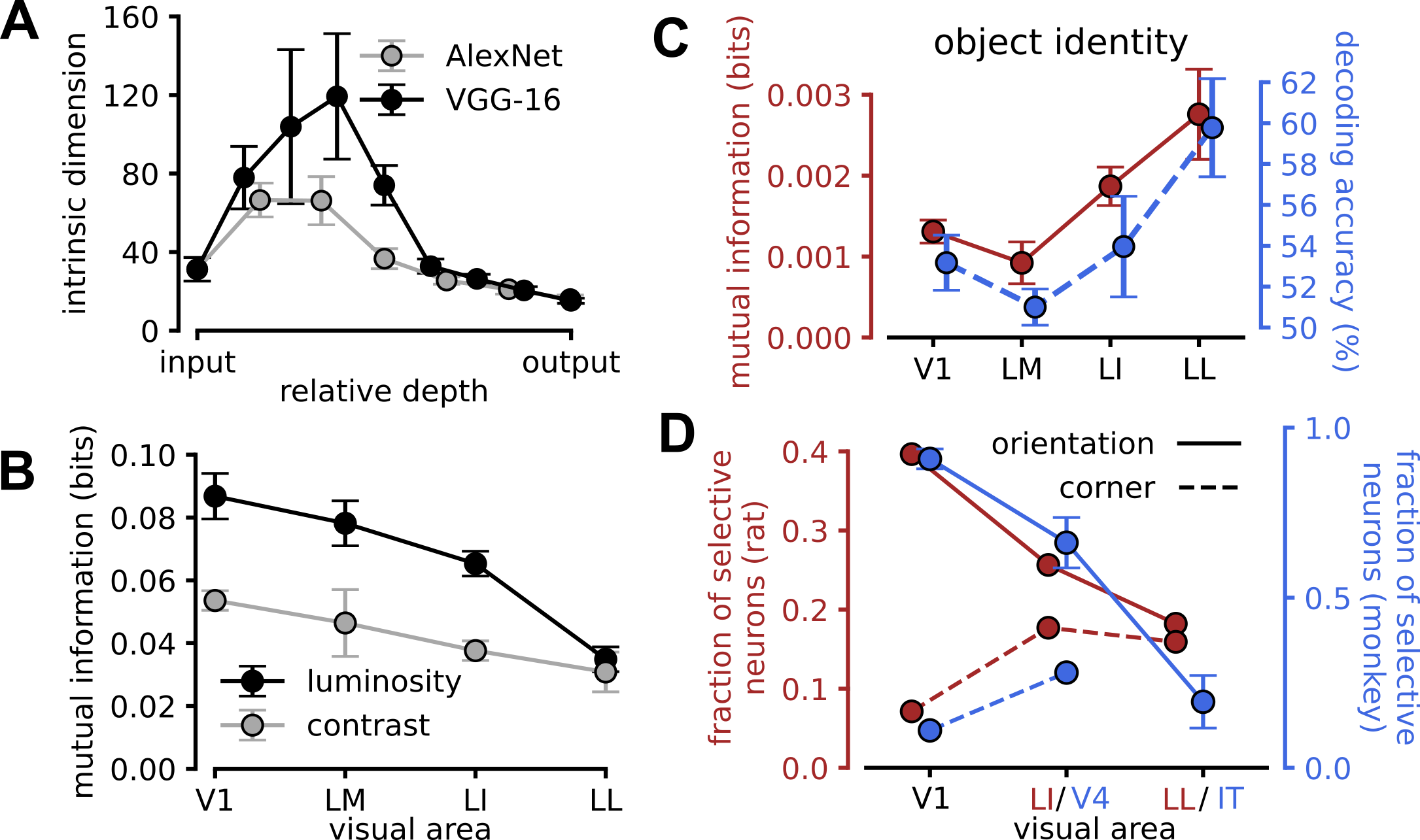}
    \caption{\textbf{Summary of previous results in CNNs and visual cortex} (\textbf{A}) Intrinsic Dimension (ID) of object representations in two deep CNNs, plotted as a function of relative depth. The error bars are standard deviations of the ID across repeated estimates (reproduced from \cite{ansuini2019intrinsic}).
    (\textbf{B})  Median information (± SE) conveyed by  single neurons recorded along the rat ventral stream (i.e.,  areas V1, LM, LI and LL) about  luminosity and contrast of  visual objects (reproduced from \cite{tafazoli2017emergence}).
    (\textbf{C}) Red curve: median information (± SE) conveyed by  single neurons in the four  areas about the identity of visual objects, each presented across a variety of distinct views. Blue curve: performance of  binary linear classifiers that were trained to discriminate two visual objects (each presented across a set of different views), based on the responses evoked by the objects in the neuronal population recorded in a given area (performance was measured over a set of validation views). Error bars are SE computed over different object pairs.  Both panels are reproduced from \cite{tafazoli2017emergence}).
    (\textbf{D}) Fraction of orientation tuned neurons (solid lines) and corner tuned neurons (dashed lines) in three different areas of the rat (red: V1, LI and LL) and monkey (blue: V1, V4 and IT) ventral stream (reproduced from \cite{matteucci2019nonlinear} and \cite{david2006spectral}).}
    \label{FIG:Figure_0}
\end{figure}

In this study, we investigated whether a similar reformatting of image information takes place along rat visual cortex and a representative CNN (VGG-16). We started from the observation that the Intrinsic Dimension (ID) of object representations follows a characteristic hunchback profile across the layers of deep networks, with an initial expansion phase, followed by a strong contraction \cite{ansuini2019intrinsic} (see examples in Figure \ref{FIG:Figure_0}A). This trend is so systematic across network architectures to raise the question of what information processing goals the two phases underlie. A possibility is that the initial expansion reflects the need of removing gradients of task-irrelevant, low-level features (e.g., luminosity and contrast) that are present in the visual input, while the  decrease underlies a gradual reformatting of the data in pursuit of a representation better suited for the classification task \cite{ansuini2019intrinsic, doimo2020hierarchical}. This idea is reminiscent of the pruning of luminosity and contrast information, and the increase of  invariance of object representations that takes place along the rat homolog of the primate ventral visual pathway \cite{tafazoli2017emergence} (Figures \ref{FIG:Figure_0}B and \ref{FIG:Figure_0}C). This parallel suggests an intriguing similarity at the level of core data-reformatting processes between artificial and natural visual hierarchies. However, it is unclear if the dimensionality of object representations follows a hunchback trend in rat visual cortex, and, conversely, if luminosity and contrast information is actively discarded across CNN layers. Moreover, other (middle-level) tuning properties exist that follow characteristic trends of variation along both the monkey and rat ventral streams \cite{matteucci2019nonlinear}. These are: (1) the fraction of neurons tuned for orientation, which decreases from primary visual cortex (V1) to higher-order areas (Figure \ref{FIG:Figure_0}D solid lines); and (2) the fraction of neurons tuned for multiple orientations (Figure \ref{FIG:Figure_0}D, dashed lines), a property thought to reflect the ability to encode corners \cite{david2006spectral}, which instead increases from V1 to downstream areas. A few studies have reported similar trends in deep networks when probed with oriented gratings \cite{hong2016explicit, matteucci2019nonlinear, benucci2022motor}, although we still lack a general assessment of how the encoding of orientation and corner information found in natural images evolve across the layers of CNNs.

The goal of our study is to understand which of these image reformatting trends found in either CNNs or visual cortex are also a signature of information processing in the other system. Specifically,  we first analysed neuronal recordings from \cite{tafazoli2017emergence} to measure the ID of object representations across the rat ventral stream, finding also in the rat the two distinct expansion-contraction phases first described in \cite{ansuini2019intrinsic} for CNNs. We then measured the  information encoded by single units in VGG-16 about a variety  of visual properties  of increasing complexity, finding that  training the network both actively distills and  prunes low- to middle-level image information, in agreement with biological observations. Finally, we tracked the evolution  of  information about object identity at both the single-unit and population level across VGG-16 layers,  exposing how such high-level information emerges sharply in late layers, again in agreement with biological findings and previous analyses of CNNs \cite{doimo2020hierarchical}.  

\section{Methods}
\label{SEC:Methods}

\subsection{Analysis of neural data}
\label{SUBSEC:Analysis_of_neural_data}

To measure how the intrinsic dimension of object representations evolves across a visual cortical processing hierarchy we analyzed the dataset recorded by \cite{tafazoli2017emergence}. These data consist of extracellular neuronal responses sampled from four visual cortical areas (228 units from V1, 131 from LM, 260 from LI, and 152 from LL), while anesthetized rats were presented with a battery of 380 stimulus conditions – i.e., 38 different views (obtained by scaling, translation, rotation, etc.) of 10 visual objects. As summarized in Figure \ref{FIG:Figure_0}A-C, object representations along this pathway were found to encode stimulus information in a way that is consistent with the existence of a functional object processing hierarchy. In our study, we computed the response of each recorded unit to every stimulus as the average number of spikes fired by the neuron across repeated presentation of the stimulus within a neuron-specific spike-count window, as defined in \cite{tafazoli2017emergence}. Each stimulus condition could thus be represented by a neuronal population vector, whose components were the responses of the neurons recorded in a given area to that stimulus. The cloud of population vectors associated to the whole set of 380 object conditions formed a data manifold, whose intrinsic dimension was measured using the nonlinear estimator defined in \cite{facco2017estimating} and previously applied to the analysis of CNNs in \cite{ansuini2019intrinsic}.

\subsection{Dataset, network architecture and estimation of the mutual information between unit activation and image properties}
\label{SUBSEC:Dataset_network_architecture}

We studied the behavior of the PyTorch implementation of VGG-16 \cite{simonyan2014very}, either randomly initialized or pre-trained on the full ImageNet dataset, with the goal of understanding how different image properties were encoded by individual units of the network as the result of training. We selected a random sub-population of $250$ units from each convolutional layer (before the ReLU activation) and from the final fully-connected layers, and recorded their activations when exposed to $1500$ input images taken from the ILSVRC2012 ImageNet validation dataset. 
 Inspired by the approach applied by \cite{tafazoli2017emergence} to  study  rat visual cortex (see Figure \ref{FIG:Figure_0}A-C), we computed Shannon mutual information $I_i^\ell \left( X_i^\ell ; Y_i^\ell \right)$ between the activation $Y_i^\ell$ of the $i$-th unit in layer $\ell$ (referred to as $u_i^\ell$ in what follows) and a given image feature $X_i^\ell$ (e.g., luminosity or contrast). The network architecture imposes for each unit $u_i^\ell$ a receptive field (RF), namely a sub-patch (denoted as $\mathsf{Image}_{\mathrm{RF}}$) of the whole image that the unit processes. As detailed in the next sections, the feature metric $X_i^\ell$ is computed by applying a specific function $\mathsf{feat}$ that maps $\mathsf{Image}_{\mathrm{RF}}$ to a real number (e.g., the luminosity intensity in the image patch) or a combination of real numbers (e.g., the two main orientations in the image patch), i.e., $\mathsf{feat} : \mathsf{Image}_{\mathrm{RF}} \to \mathbb{R}$ or $\mathsf{feat} : \mathsf{Image}_{\mathrm{RF}} \to \mathbb{R}^2$.
 
Given unit $u_i^\ell$, the values taken by its activation $Y_i^\ell$ over the set of input images yield a unit-specific activation distribution $p_Y ( y )$ (for simplicity, we dropped the unit and layer indexes). Similarly, the values taken by the feature metric $X_i^\ell$ yield a unit-specific (i.e., RF-specific) distribution $p_X \left(x \right)$ (e.g., of luminosity intensity levels). In our experiments, both distributions were discretized into $20$ equi-spaced bins. By computing the joint distribution of activation and feature values $p_{X,Y} \left(x, y \right)$, we estimated, for each unit, the mutual information between activation and feature metric:
 
 \begin{equation}
     \ I \left(X ; Y \right) = \sum_{x \in \mathcal{X}, y \in \mathcal{Y}} p_{X,Y} \left(x, y \right) \mathrm{log} \frac{p_{X,Y} \left(x, y \right)}{p_X \left(x \right) p_Y \left( y \right)}.
\label{EQ:Information}
\end{equation}

 To allow for a better comparison among the various layers and the different image features used in our analysis, the mutual information was normalized by the entropy of the distribution of the feature metric $H \left( X \right) = -\sum_{x \in \mathcal{X}} p_X \left( x \right) \mathrm{log} p_X \left( x \right)$. The final estimate of the information conveyed by the units of a given layer about the feature metric was computed as $U^\ell \left( X | Y \right) = \mathbb{E}_i \left[ I_i^\ell \left(X_i^\ell ; Y_i^\ell \right)/H_i^\ell \left( X_i^\ell \right) \right]$, where $\mathbb{E}_i$ is the expected value over all units $i$ of layer $\ell$. Importantly, although such unit-averaging was performed on a sub-population of $\mathcal{O} \left( 10^2 \right)$ units, the variability of $U^\ell $ across independent experiment realizations (different units and stimuli) was very low, as shown by the error bars reported in Figures \ref{FIG:Figure_2} and \ref{FIG:Figure_4}. The limited sampling bias for the mutual information was corrected with the the Panzeri-Treves method \cite{panzeri1996analytical, panzeri2007correcting}.

\subsection{Definition of the metrics to quantify visual features}
\label{SUBSEC:Definition_of_the_metrics}

In our analysis, each image patch $\mathsf{Image}_{\mathrm{RF}}$ falling within the RF of a unit was quantified by an array of four different visual properties of increasing complexity: 1)  luminosity; 2)  contrast; 3) orientation of the dominant edge (if any); and 4) orientations of the two dominant edges (if any), which is a proxy for the orientation and width of the dominant corner. Therefore $\mathsf{feat} \in \left\{ \mathsf{luminosity}, \mathsf{contrast}, \mathsf{orientation}, \mathsf{corner} \right\}$. 

Luminosity can be easily defined as the average pixel-intensity in the image path: $\mathsf{luminosity} = \mathsf{mean} \left( \mathsf{Image}_\mathrm{RF} \right)$. Contrast quantifies the amount of luminosity variation  in the patch and was computed   as: $\mathsf{contrast} = \mathsf{mean} \left( \mathsf{Sobel} \ast \mathsf{Image}_\mathrm{RF} \right)$, where $\mathsf{Sobel}$ denotes the Sobel kernel and $\ast$ is the convolution operator (the Sobel transform is a standard approach to compute image gradients \cite{gonzalez2008digital}).


    
    
    
    
    

The dominant orientation of in an image patch is less straightforward to quantify, because  of the large variation in RF size across the layers of the network and the complexity of the natural scenes in ImageNet. At very low resolution, such as for individual units in early layers in VGG-16 (which have $3\times 3$ RF size), no meaningful orientation can be computed. For units in late layers, which process the entire scene, multiple prominent orientations might coexist or not exist at all. More generally, image patches span a spectrum of scene orientation strength, ranging from those containing one or more sharp edges to those featuring none. To deal with such variability, we developed a two-stage, compute-and-filter approach. The $\mathsf{orientation}$ estimation routine is based on  Fourier Analysis  and defines the dominant orientation of the  patch $\theta^\star$ as the angle of highest power of its Fourier spectrum (see Algorithm 1 in the Supplementary Material for a detailed pseudo-code of the pipeline). In addition, the function provides  an orientation strength index  $\xi \in \left[0, 1 \right]$, which peaks for images containing at least one very sharp edge. Before measuring orientation information, we  ranked the pool  of sampled units in each layer by computing, for each unit, the average of the orientation strength index $\xi$ across the full set of $1500$ input images. Out of the initial population of $250$ units we only retained the  $200$ units with the largest average index. In addition, for each selected unit, we only considered the  $500$ images with the largest index $\xi$.

The corner  feature was quantified  as the pair of orientations  of the two most prominent edges in the image patch.  Specifically, the $\mathsf{corner}$ estimation routine applies  Fourier analysis and peak-finding algorithms to identify the two dominant orientations  $\theta^\star_1$ and $\theta^\star_2$ in a patch, along with a corner strength index $\zeta \in \left[0, 1 \right]$ \cite{david2006spectral, matteucci2019nonlinear}, which is large when at least two orientations with similar  power are detected in the Fourier spectrum, while it becomes negligible both for no-peak and single-peaked angular spectra (see Algorithm 2 in the Supplementary Materials for a pseudo-code description of the complete pipeline). We used this index following the same rationale as for the $\xi$ index of orientation strength, this time ranking units and input images based on their corner strength $\zeta$  and retaining a population of $200$ units, each tested with a sample of $400$ images.

\section{Results}
\label{SEC:Results}

\subsection{Intrinsic dimension of object representations along the rat ventral stream}
\label{SUBSEC:Intrinsic_dimension_of_object}

\begin{figure}[t]
    \centering
    \includegraphics[width=11cm]{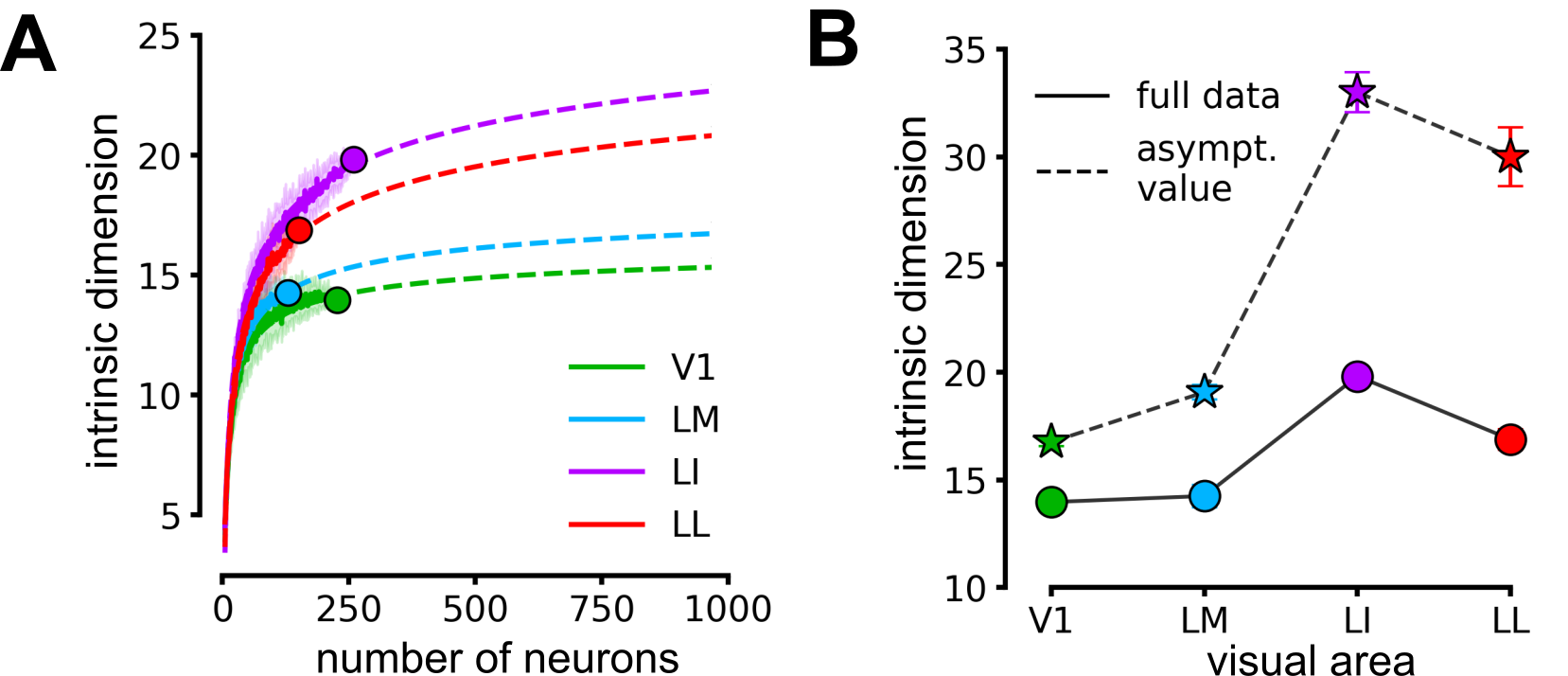}
    \caption{\textbf{Intrinsic Dimension of neural representations} 
    (\textbf{A}) Estimation of the ID as a function of the number of neurons considered in the four visual areas (solid lines). Shadings correspond to the SD of multiple estimates with randomly sampled neuronal sub-populations, while  circles mark the estimates obtained with the full populations in each area. Dashed lines are power-law fits to the data. (\textbf{B}) The ID estimates obtained for the full populations (circles) and for the asymptotic values of the fits (stars) are plotted as a function of the rank of the areas along the rat ventral stream. Error bars  are the standards deviation of the values returned by the fit.}
    \label{FIG:Figure_1}
\end{figure}

We applied the nonlinear ID estimator Two-NN \cite{facco2017estimating} to compute the intrinsic dimension of object representations in  four visual cortical areas (V1, LM, LI and LL) of the rat ventral stream,  as a function of the number of units included in the population vector space (Figure \ref{FIG:Figure_1}A, solid lines), up to the maximal number of units available in each area (circles). In addition,  we extracted the asymptotic values of the ID (stars in Figure \ref{FIG:Figure_1}B) via  power-law fits (dashed lines) to control for finite-size effects. At any population size considered, the ranking of the visual areas in terms of the estimated ID  was remarkably stable, with  V1  featuring the lowest ID, LI the highest, and LM and LL reaching intermediate values. More importantly, plotting the ID in each area as a function of its rank along the cortical processing hierarchy (Figure \ref{FIG:Figure_1}B) revealed a characteristic "hunchback" profile, with an initial expansion (from V1 to LI), followed by a contraction (from LI to LL). This trend is consistent with the one  observed in deep networks (see Figure \ref{FIG:Figure_0}A) by \cite{ansuini2019intrinsic}, who conjectured that the initial ID expansion  was due to the pruning of low-level image information (e.g., luminosity and contrast). Our result strongly supports this intuition, since the alternation of the expansion-contraction phases is now observed along an object processing pathway where such pruning has been shown to  take place (see Figure \ref{FIG:Figure_0}B) \cite{tafazoli2017emergence}.

\subsection{Encoding of low- to middle-level visual features in single units of VGG-16}
\label{SUBSEC:Encoding_of_low_to_middle_level}

We now turn to  the other question addressed in our study, namely investigating if the information about low-level image features is  actively discarded in artificial networks in a manner that resembles the one observed in rats. 
Having defined a set of metrics to quantify image features of increasing complexity (see Section \ref{SUBSEC:Definition_of_the_metrics}), we measured how much information about these features was encoded by the  activation of individual units across the layers of VGG-16 (see Section \ref{SUBSEC:Dataset_network_architecture}). 

\begin{figure}[t]
    \centering
    \includegraphics[width=11cm]{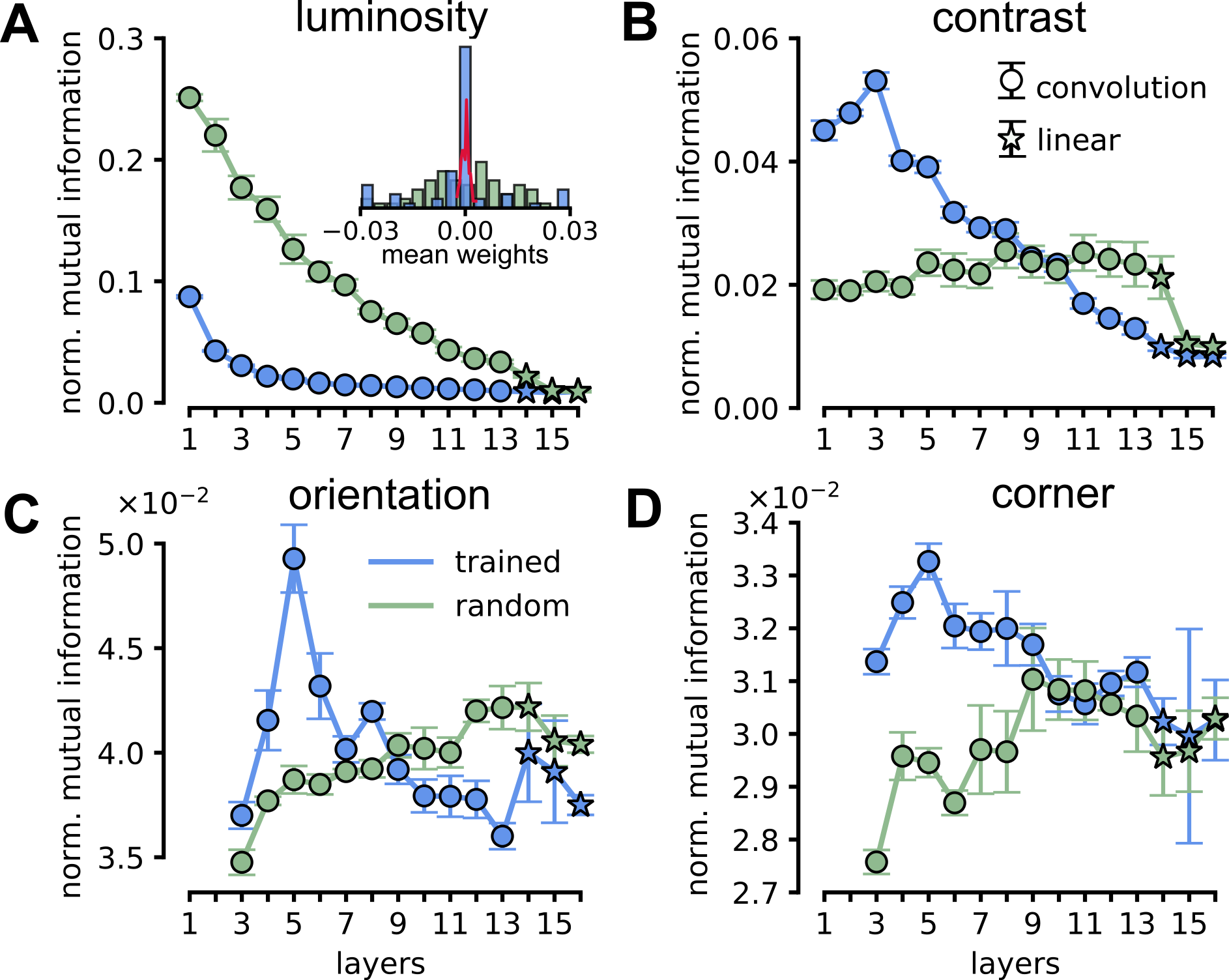}
    \caption{\textbf{Distillation and pruning of image information in VGG-16} (\textbf{A}) Mean normalized  information conveyed by  VGG-16 units about image luminosity for a trained (blue) and random (green) network. Error bars are standard deviations over five realizations of the experiment (independent sampling of units, images and random weights). Circles represent convolutional layers, while stars indicate fully connected layers. Inset: distribution of the average weights of the units in the first layer of the trained and random network (blue and green bars) and in the forth convolutional layer of the latter. (\textbf{B}-\textbf{D}) Same as in \textbf{A}, but for the information conveyed by VGG-16 units about image contrast, orientation and corners (i.e., joint orientation of two prominent edges).}
    \label{FIG:Figure_2}
\end{figure}

We found that information about luminosity was a monotonic decreasing function of the layer's depth, with training producing a very large luminosity information loss in the very first layer (compare blue and green curves in Figure \ref{FIG:Figure_2}A). Intuitively, this  can be  explained by the fact that learning spatially structured convolutional kernels will tend to produce both positive and negative weights with balanced, near-zero average, which are poorly sensitive to the mean luminosity within a unit's RF. By contrast, randomly assigned weights will often  have  the same sign, at least for the small kernels of the early layers, yielding activations that are proportional to the luminous energy falling within a unit's RF. This intuition was confirmed by comparing the distributions of the average weights for the $3\times3$ kernels of the first layer in the trained and untrained network (bar plot in the inset). The gradual monotonic decrease that was nevertheless observed in the untrained network is explained by the fact that randomly assigned weights, in case of increasingly larger kernels, will progressively tend to the zero-average condition (inset, red line).

If training produces spatially structured kernels, units in early layers should not only lose sensitivity to luminosity, but also become sensitive to image contrast. 
Our mutual information analysis confirmed this intuition, showing that the units of the initial layers encoded a larger amount of contrast information in the trained network, as compared to the untrained one (Figure \ref{FIG:Figure_2}B, blue vs. green curve). In addition, as a result of training, contrast information grew steadily in the early convolutional layers, reaching a peak in the third one, but then decayed sharply in the following layers, eventually attaining values that were lower than those of the untrained network. This suggests that learning representations that are useful to process and classify natural images requires to  first distill contrast information in the units of early layers, followed by actively discarding such information in  later processing stages. The pruning of contrast and luminosity information matches the results in rat visual cortex \cite{tafazoli2017emergence} (see Figure \ref{FIG:Figure_0}B). We note that the analysis of rat data did not reveal the initial rise of contrast information found in VGG-16, but this is unsurprising, given that the rat dataset did not contain recordings from the processing stages that precede V1 (i.e., retina and thalamus), which would correspond to VGG-16 very first layers.

We next considered visual features of increasing complexity,  measuring the amount of information encoded by VGG-16 units about the dominant orientations of the image patches falling within  their RFs. This analysis was applied  only to units for which enough input images existed that  contained, in the  patch falling within the units' RFs,  a sufficiently prominent oriented edge (see Section 2.3).  Moreover,  we excluded from the analysis the first two  layers of the network, because their units have RFs that are too small for the $\mathsf{orientation}$ estimate to be meaningful. When visualized as a function of layer depth,  orientation information in the trained network followed a hunchback profile (Figure \ref{FIG:Figure_2}C, blue curve), raising sharply and reaching a peak in the fifth convolutional layer, i.e., at a later stage than contrast information (see Figure \ref{FIG:Figure_2}B), consistently with the hierarchically higher nature of the orientation feature. Following the peak, orientation information dropped  sharply in the deeper layers. As for luminosity and contrast, this trend was the result of training, as it was not observed in the randomly initialized network (green curve). And again, as for contrast, orientation information, once distilled in individual units of early layers, was then actively discarded in the following processing stages, becoming lower than for the untrained network - a finding consistent with the loss of orientation tuning found along the ventral stream \cite{matteucci2019nonlinear} (see Figure \ref{FIG:Figure_0}D, solid lines). Just like for contrast, no initial rise of orientation tuning was observed along the rat ventral stream, because no data were available from subcortical areas where orientation tuning is known to be much less prominent \cite{durand2016comparison}.

Finally, considering  features of  even greater complexity, we measured the information conveyed by VGG-16 units about the joint orientation of two dominant edges, i.e., the corner information (again, this analysis was applied only to cases where the image patch falling within a unit RF contained  a sufficiently prominent corner; see Section 2.3). As for orientation, also corner information  varied across the layers of the trained network according to a hunchback profile (Figure \ref{FIG:Figure_2}D, blue curve), again peaking in the fifth convolutional layer, and again being discarded in  deeper processing stages, but  more gradually than orientation information, reaching a sort of plateau in middle layers. Once more, this trend was not observed in the untrained network (green curve) and was instead consistent with the increase of neurons tuned for pairs of orientations found along the  ventral stream \cite{matteucci2019nonlinear} (see Figure \ref{FIG:Figure_0}D, dashed lines).       

\subsection{Effective pruning of low-level information requires training}
\label{SUBSEC:Effective_pruning_of_low_level_information}

\begin{figure}[htb]
    \centering
    \includegraphics[width=12cm]{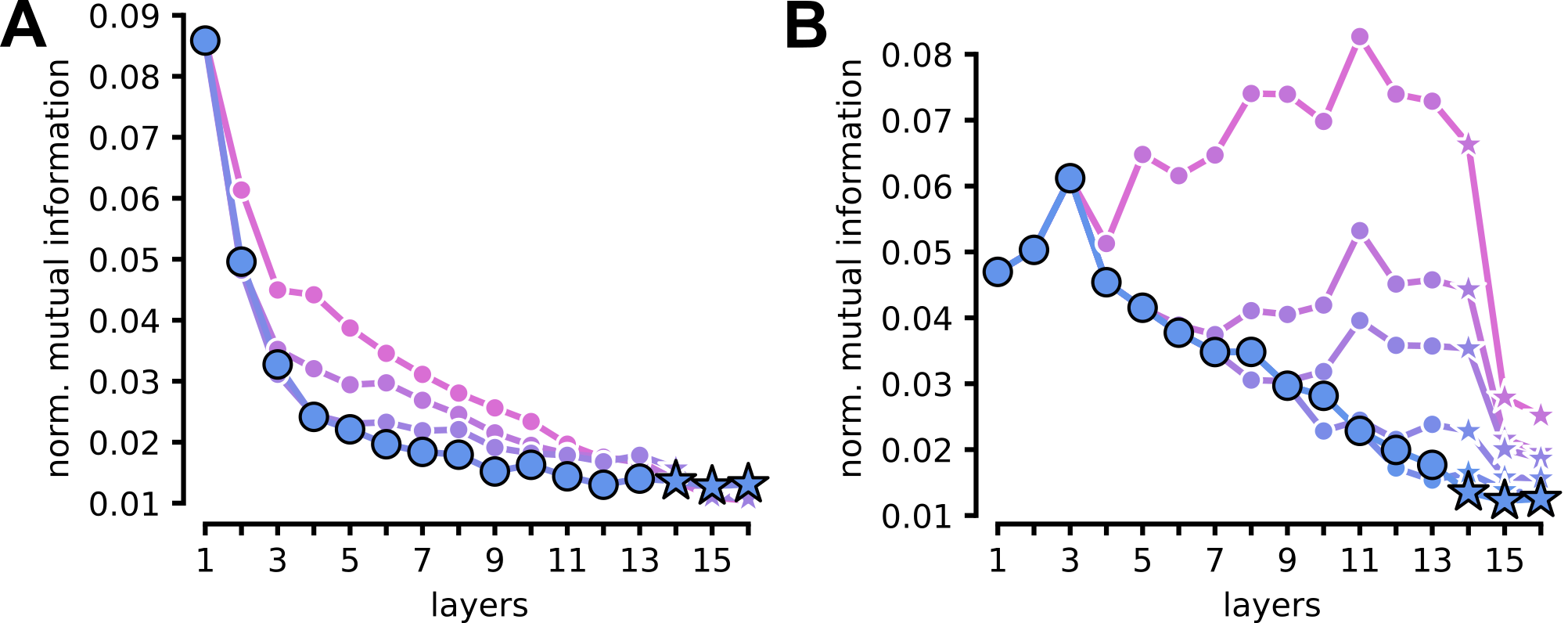}
    \caption{\textbf{Training results in active pruning of low-level image information} (\textbf{A}) Mean normalized  information conveyed by  VGG-16 units about image luminosity for a trained network (thick blue line; same curve as in Figure \ref{FIG:Figure_2}A) and for three additional hybrid network configurations (purple lines) that have been trained only until layer $\ell^\star$ (with weights in the following layers having been left randomly initialized). The gradients of purple (from pink to violet) correspond to progressively larger $\ell^\star$ values, i.e., $\ell^\star \in \left\{1, 2, 3 \right\}$. (\textbf{B}) Same as in \textbf{A}, but for the information conveyed by VGG-16 units about image contrast. Here the thick blue line is the same curve as in Figure \ref{FIG:Figure_2}B and  $\ell^\star \in \left\{3, 5, 7, 9, 11 \right\}$.}
    \label{FIG:Figure_3}
\end{figure}

One of the most intriguing findings of our experiments is that training is necessary not only to build sensitivity for low- and middle-level visual features, but also plays the complementary role of pruning this information, once it has been distilled in individual units of early layers.  To better understand the extent to which   information pruning is actively enforced by training, we considered a hybrid VGG-16 network constructed as follows: layers $\ell \le \ell^\star$ shared the same weights as the fully-trained (on ImageNet) VGG-16, while weights in layers $\ell > \ell^\star$ were left randomly initialized. By letting $\ell^\star$ vary, one could visualize the effect of  random transformations after a given checkpoint ($\ell^\star$) and ask whether the observed decay of feature information (see Figure \ref{FIG:Figure_2}A and \ref{FIG:Figure_2}B as examples) is a direct consequence of training (active information pruning) or is merely the result of architectural constraints. We found that training played an active role in pruning luminosity information (Figure \ref{FIG:Figure_3}A), with the information profile of the fully-trained network (blue curve) being consistently  lower  with respect to the profiles  obtained for hybrid networks with intermediate $\ell^\star$ checkpoints (purple curves). The effect of training was even more striking for contrast information (see Figure \ref{FIG:Figure_3}B), which, in the hybrid networks, displayed a growing trend through the random convolutional layers, before finally dropping in the  second fully-connected layer.

\subsection{Information on object identity emerges in late layers at both the single unit and population level}
\label{SUBSEC:Information_on_object_identity_emerges_lately}

\begin{figure}[b]
    \centering
    \includegraphics[width=\textwidth]{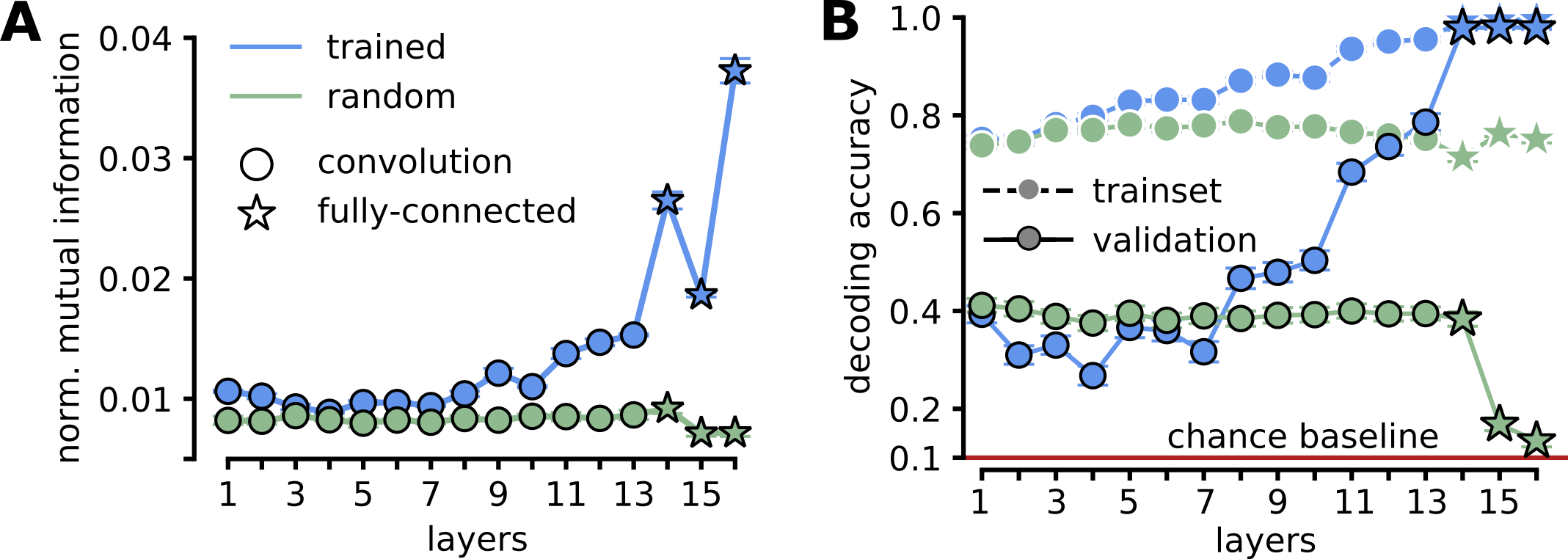}
    \caption{\textbf{Evolution of category information across VGG-16 layers} (\textbf{A}) Mean normalized  information conveyed by  VGG-16 units about image category for a trained (blue line) and a random   (green line) VGG-16 network. As in Figure \ref{FIG:Figure_2}, error bars are standard deviations over five realizations of the experiment. (\textbf{B}) Training and validation performance (dashed and solid lines respectively) of linear SVM classifiers that were trained to  predict the labels of images belonging to $10$ selected Imagenet categories ($250$ and $50$ images per category were used, respectively, for training and validation), based on the activations of a pool of $250$ VGG-16 units sampled from each layer. Data points are averages (± SD) over $200$ sub-populations of $100$ units that were randomly sampled from each pool of $250$ units.}
    \label{FIG:Figure_4}
\end{figure}

The  VGG-16 network used in our experiments was  pretrained to achieve high  classification performance on Imagenet. Thus, all the information about the low- to middle-level visual features explored in our analyses must have been harvested (and then pruned) by the network in the attempt to maximize the separability  of image categories in its output layer. Having reported how information about several such features peaked in early convolutional layers, we next asked how information about image category evolved across the network. Intuition suggests that it should peak in the very last layer, where readout takes place. It is however unclear how such information varies along the network depth, especially the one encoded by individual units. In the rat ventral stream, information about object category encoded by single neurons has been found to rise from low to high visual areas, with a matching increase in the ability of neuronal populations to support invariant recognition \cite{tafazoli2017emergence} (see Figure \ref{FIG:Figure_0}C). In \cite{doimo2020hierarchical}, using a neighbourhood regularity metric, it was shown how representation-support for image category emerged sharply in late layers of various CNN architectures. 

Here, we  measured the category information encoded by VGG-16 units by using the label of the $1500$ test images as the feature variable $X_i^\ell$ in Eq. \eqref{EQ:Information}. We found that this metric remained low and stable for about half the depth of the trained network, increasing smoothly in the last convolutional layers and then abruptly in the fully connected ones (Figure \ref{FIG:Figure_4}A, blue curve), while no trend was observed for the random network (green curve). This result resonates with that in \cite{doimo2020hierarchical}, again indicating a late and sharp rise in image category information.

Next, we investigated how easily accessible was such category information  encoded by single units. To this aim, we trained   linear SVMs  to predict the image labels based on the activity of a  pool of 250 units in a layer (Supplementary Material, section B). We found a  growth of decoding accuracy  (Figure \ref{FIG:Figure_4}B, blue curve) that  tracked the  increase of category information observed, at the  single unit level, in the second half of the convolutional layers (compare to Figure \ref{FIG:Figure_4}A). This suggests that, similarly to what observed along  the  rat ventral stream, the concentration of category information in single units plays a role in supporting the linear readout of category label at the population level. As expected, the decoding accuracy matched the expected close-to-perfect performance in the final, fully connected layers (see Figure \ref{FIG:Figure_4}B, blue lines). Interestingly, even the random network supported above-chance decoding beyond the training domain (solid green curve), with both the trained and random configurations remaining closely tied for half of the computation depth (mirroring, again, what observed for the category information at the single unit level).

\section{Conclusions and Discussion}

The visual ventral stream  \cite{dicarlo2012does} and deep convolutional neural networks \cite{lecun2015deep} are very effective solutions to the problem of object vision. The precise extent to which these two classes of systems process visual information based on similar principles is an open question and an active field of investigation \cite{kriegeskorte2015deep,yamins2016using,lindsay2021convolutional}. Here, we compared  the rat ventral stream and a widely used CNN (VGG-16)  by tracking how image representations are progressively reformatted across the two  hierarchies. Differently from previous studies, our goal was not to use CNNs as models to predict the tuning of visual cortical neurons or the spatial structure of their RFs \cite{yamins2014performance, yamins2016using,ponce2019evolving,bashivan2019neural,walker2019inception}. Rather, we took inspiration from  recent studies showing that some key statistical properties of image representations follow very specific trends of variation across either the artificial or the cortical processing hierarchies \cite{tafazoli2017emergence,ansuini2019intrinsic}, and we looked for them in the other system.

In this study, we focused on the rat brain as the biological term of our comparison with CNNs, rather than on the primate brain, which certainly possesses a more advanced visual system. This is because, despite the long tradition of vision neuroscience in primates, many of the information processing trends we explored in this work have been systematically measured only in the rat \cite{tafazoli2017emergence}. This is related to the fact that, in monkeys, it is difficult to record from the whole ventral stream using the same battery of visual stimuli. Typically, no more than a pair of cortical areas are investigated in a single study (e.g., V4 and IT \cite{rust2010selectivity}) and object representations in V1 are often simulated rather than measured, when compared to higher-order areas \cite{yamins2014performance,hong2016explicit}. By contrast, in rats it is possible to probe with the same stimuli V1 and the whole progression of lateral extrastriate areas (LM, LI and LL) that play the functional role of an object-processing pathway \cite{vermaercke2014functional,tafazoli2017emergence,piasini2021temporal}. This makes it possible to analyze  a cortical hierarchy that is deep enough for a meaningful comparison with a CNN. 

Our analyses yielded three main results. First, the ID of object representations across the rat ventral stream varied according to the same hunchback profile previously found by \cite{ansuini2019intrinsic} in CNNs (compare Figure \ref{FIG:Figure_1}B to Figure \ref{FIG:Figure_0}A). Here it should be noted that, compared to the dramatic ID contraction observed in the final layers of CNNs, the drop found from LI to LL was much smaller. This is not surprising, because LL, despite its high rank along the rat visual cortical progression, is not the final stage of the   hierarchy. In fact, the rat ventral stream possesses at least one higher-order area (TO, \cite{vermaercke2014functional,vinken2016neural}; not probed in \cite{tafazoli2017emergence}). Additionally, the deepest layers of CNNs contain representations that are highly specialized for the  classification tasks they were trained on, suggesting a better match with cortical regions involved in memory and decision making (such as perirhinal, posterior parietal and prefrontal cortex) than with purely  sensory areas. It is in these regions that representations may be expected to become as low dimensional as those found in CNNs' final layers \cite{orlandi2021distributed,brincat2018gradual}.

Our second main result is that  information about  low-level  features (luminosity and contrast) encoded by individual units of VGG-16 was progressively pruned across the processing hierarchy, as previously found by \cite{tafazoli2017emergence} across rat visual cortical areas (compare Figure \ref{FIG:Figure_2}A-B to Figure \ref{FIG:Figure_0}B). A similar pruning was observed also in the case of orientation and corner information (Figure \ref{FIG:Figure_2}C-D), with the difference that sensitivity to these higher-order features started low and had first to be progressively distilled through processing along the first  layers (a trend that, although less prominent, was also observed for contrast information). These trends are consistent with the drop of orientation tuning and the increase of tuning for multiple orientations (corners) found in both the rat and monkey ventral streams \cite{matteucci2019nonlinear} (Figure \ref{FIG:Figure_0}D). In addition, they are consistent with the orientation tuning profiles reported across CNN layers by a few studies  \cite{hong2016explicit, matteucci2019nonlinear, benucci2022motor} and with the way the spatial structure of convolutional kernels evolves across CNN layers, where early Gabor-like kernels are replaced by filters with more complex geometries in late layers \cite{zeiler2014visualizing}. Importantly, our analyses clearly show that information pruning is not a trivial byproduct of architectural constraints (e.g., RFs becoming larger as a function of layer depth), but  is an active process that takes place, across the whole network, as the result of training (Figure \ref{FIG:Figure_2} and \ref{FIG:Figure_3}). This suggests that, in  hierarchical visual processing systems (biological and artificial alike), sensitivity to features that are required for the buildup of higher abstractions (e.g., contrast for edges; edges for corners; corners for shapes; etc..) might  become useless or even harmful for further learning in deep layers, and is consequently discarded.

Finally, our experiments revealed that the growth of  classification accuracy afforded by image representations across VGG-16 layers closely tracked the increase of category information encoded by individual units (Figure \ref{FIG:Figure_4}). This result is remarkably consistent with the tight relationship found, in the rat ventral stream, between the view-invariant object information encoded by single neurons and the power of neuronal populations to support invariant object recognition (see Figure \ref{FIG:Figure_0}C). Overall, this suggests that low/middle-level image information and higher-order categorical information trade off along visual processing hierarchies, echoing previous observations that have emphasized the role of learning in suppressing irrelevant information \cite{shwartz-zivOpeningBlackBox2017a}. 

Taken together, these findings point to the existence of  a functional relationship between dimensionality expansion/reduction of object representations and distillation/pruning of various kinds of image information, suggesting that such relationship is likely a fundamental  property of both biological and artificial visual processing architectures.

\begin{ack}
We thank A. Ansuini for his help on getting started with the computation of intrinsic dimension in deep nets and neuronal data. We thank D. Doimo and L. Porta for suggestions on the implementations of our analyses. We thank A. Benucci  for his feedback on the interpretation of our findings.

This work was supported by a European Research Council Consolidator Grant (project no. 616803-LEARN2SEE to D.Z)

\end{ack}

\printbibliography

@misc{shwartz-zivOpeningBlackBox2017a,
  title = {Opening the {{Black Box}} of {{Deep Neural Networks}} via {{Information}}},
  author = {{Shwartz-Ziv}, Ravid and Tishby, Naftali},
  year = {2017},
  month = apr,
  number = {arXiv:1703.00810},
  eprint = {1703.00810},
  eprinttype = {arxiv},
  primaryclass = {cs},
  institution = {{arXiv}},
  doi = {10.48550/arXiv.1703.00810},
  archiveprefix = {arXiv}
}

@article{facco2017estimating,
  title={Estimating the intrinsic dimension of datasets by a minimal neighborhood information},
  author={Facco, Elena and d’Errico, Maria and Rodriguez, Alex and Laio, Alessandro},
  journal={Scientific reports},
  volume={7},
  number={1},
  pages={1--8},
  year={2017},
  publisher={Nature Publishing Group}
}

@article{tafazoli2017emergence,
  title={Emergence of transformation-tolerant representations of visual objects in rat lateral extrastriate cortex},
  author={Tafazoli, Sina and Safaai, Houman and De Franceschi, Gioia and Rosselli, Federica Bianca and Vanzella, Walter and Riggi, Margherita and Buffolo, Federica and Panzeri, Stefano and Zoccolan, Davide},
  journal={Elife},
  volume={6},
  pages={e22794},
  year={2017},
  publisher={eLife Sciences Publications Limited}
}

@article{ansuini2019intrinsic,
  title={Intrinsic dimension of data representations in deep neural networks},
  author={Ansuini, Alessio and Laio, Alessandro and Macke, Jakob H and Zoccolan, Davide},
  journal={Advances in Neural Information Processing Systems},
  volume={32},
  year={2019}
}

@article{doimo2020hierarchical,
  title={Hierarchical nucleation in deep neural networks},
  author={Doimo, Diego and Glielmo, Aldo and Ansuini, Alessio and Laio, Alessandro},
  journal={Advances in Neural Information Processing Systems},
  volume={33},
  pages={7526--7536},
  year={2020}
}

@article{yamins2014performance,
  title={Performance-optimized hierarchical models predict neural responses in higher visual cortex},
  author={Yamins, Daniel LK and Hong, Ha and Cadieu, Charles F and Solomon, Ethan A and Seibert, Darren and DiCarlo, James J},
  journal={Proceedings of the national academy of sciences},
  volume={111},
  number={23},
  pages={8619--8624},
  year={2014},
  publisher={National Acad Sciences}
}

@article{yamins2016using,
  title={Using goal-driven deep learning models to understand sensory cortex},
  author={Yamins, Daniel LK and DiCarlo, James J},
  journal={Nature neuroscience},
  volume={19},
  number={3},
  pages={356--365},
  year={2016},
  publisher={Nature Publishing Group}
}

@article{richards2019deep,
  title={A deep learning framework for neuroscience},
  author={Richards, Blake A and Lillicrap, Timothy P and Beaudoin, Philippe and Bengio, Yoshua and Bogacz, Rafal and Christensen, Amelia and Clopath, Claudia and Costa, Rui Ponte and de Berker, Archy and Ganguli, Surya and others},
  journal={Nature neuroscience},
  volume={22},
  number={11},
  pages={1761--1770},
  year={2019},
  publisher={Nature Publishing Group}
}

@article{bashivan2019neural,
  title={Neural population control via deep image synthesis},
  author={Bashivan, Pouya and Kar, Kohitij and DiCarlo, James J},
  journal={Science},
  volume={364},
  number={6439},
  pages={eaav9436},
  year={2019},
  publisher={American Association for the Advancement of Science}
}

@article{goodfellow2014explaining,
  title={Explaining and harnessing adversarial examples},
  author={Goodfellow, Ian J and Shlens, Jonathon and Szegedy, Christian},
  journal={arXiv preprint arXiv:1412.6572},
  year={2014}
}

@incollection{kurakin2018adversarial,
  title={Adversarial examples in the physical world},
  author={Kurakin, Alexey and Goodfellow, Ian J and Bengio, Samy},
  booktitle={Artificial intelligence safety and security},
  pages={99--112},
  year={2018},
  publisher={Chapman and Hall/CRC}
}

@inproceedings{zeiler2014visualizing,
  title={Visualizing and understanding convolutional networks},
  author={Zeiler, Matthew D and Fergus, Rob},
  booktitle={European conference on computer vision},
  pages={818--833},
  year={2014},
  organization={Springer}
}

@article{matteucci2019nonlinear,
  title={Nonlinear processing of shape information in rat lateral extrastriate cortex},
  author={Matteucci, Giulio and Marotti, Rosilari Bellacosa and Riggi, Margherita and Rosselli, Federica B and Zoccolan, Davide},
  journal={Journal of Neuroscience},
  volume={39},
  number={9},
  pages={1649--1670},
  year={2019},
  publisher={Soc Neuroscience}
}

@article{benucci2022motor,
  title={Motor-related signals support localization invariance for stable visual perception},
  author={Benucci, Andrea},
  journal={PLoS computational biology},
  volume={18},
  number={3},
  pages={e1009928},
  year={2022},
  publisher={Public Library of Science San Francisco, CA USA}
}

@article{hong2016explicit,
  title={Explicit information for category-orthogonal object properties increases along the ventral stream},
  author={Hong, Ha and Yamins, Daniel LK and Majaj, Najib J and DiCarlo, James J},
  journal={Nature neuroscience},
  volume={19},
  number={4},
  pages={613--622},
  year={2016},
  publisher={Nature Publishing Group}
}

@article{simonyan2014very,
  title={Very deep convolutional networks for large-scale image recognition},
  author={Simonyan, Karen and Zisserman, Andrew},
  journal={arXiv preprint arXiv:1409.1556},
  year={2014}
}

@article{ponce2019evolving,
  title={Evolving images for visual neurons using a deep generative network reveals coding principles and neuronal preferences},
  author={Ponce, Carlos R and Xiao, Will and Schade, Peter F and Hartmann, Till S and Kreiman, Gabriel and Livingstone, Margaret S},
  journal={Cell},
  volume={177},
  number={4},
  pages={999--1009},
  year={2019},
  publisher={Elsevier}
}

@article{li2008unsupervised,
  title={Unsupervised natural experience rapidly alters invariant object representation in visual cortex},
  author={Li, Nuo and DiCarlo, James J},
  journal={science},
  volume={321},
  number={5895},
  pages={1502--1507},
  year={2008},
  publisher={American Association for the Advancement of Science}
}

@article{matteucci2020unsupervised,
  title={Unsupervised experience with temporal continuity of the visual environment is causally involved in the development of V1 complex cells},
  author={Matteucci, Giulio and Zoccolan, Davide},
  journal={Science advances},
  volume={6},
  number={22},
  pages={eaba3742},
  year={2020},
  publisher={American Association for the Advancement of Science}
}

@article{david2006spectral,
  title={Spectral receptive field properties explain shape selectivity in area V4},
  author={David, Stephen V and Hayden, Benjamin Y and Gallant, Jack L},
  journal={Journal of neurophysiology},
  volume={96},
  number={6},
  pages={3492--3505},
  year={2006},
  publisher={American Physiological Society}
}

@article{dicarlo2012does,
  title={How does the brain solve visual object recognition?},
  author={DiCarlo, James J and Zoccolan, Davide and Rust, Nicole C},
  journal={Neuron},
  volume={73},
  number={3},
  pages={415--434},
  year={2012},
  publisher={Elsevier}
}

@article{kriegeskorte2015deep,
  title={Deep neural networks: a new framework for modeling biological vision and brain information processing},
  author={Kriegeskorte, Nikolaus},
  journal={Annual review of vision science},
  volume={1},
  pages={417--446},
  year={2015},
  publisher={Annual Reviews}
}

@article{lecun2015deep,
  title={Deep learning},
  author={LeCun, Yann and Bengio, Yoshua and Hinton, Geoffrey},
  journal={nature},
  volume={521},
  number={7553},
  pages={436--444},
  year={2015},
  publisher={Nature Publishing Group}
}

@article{lindsay2021convolutional,
  title={Convolutional neural networks as a model of the visual system: Past, present, and future},
  author={Lindsay, Grace W},
  journal={Journal of cognitive neuroscience},
  volume={33},
  number={10},
  pages={2017--2031},
  year={2021},
  publisher={MIT Press One Rogers Street, Cambridge, MA 02142-1209, USA journals-info~…}
}

@article{orlandi2021distributed,
  title={Distributed context-dependent choice information in mouse dorsal-parietal cortex},
  author={Orlandi, Javier and Adbolrahmani, Mohammad and Aoki, Ryo and Lyamzin, Dmitry and Benucci, Andrea},
  year={2021}
}

@article{piasini2021temporal,
  title={Temporal stability of stimulus representation increases along rodent visual cortical hierarchies},
  author={Piasini, Eugenio and Soltuzu, Liviu and Muratore, Paolo and Caramellino, Riccardo and Vinken, Kasper and Op de Beeck, Hans and Balasubramanian, Vijay and Zoccolan, Davide},
  journal={Nature communications},
  volume={12},
  number={1},
  pages={1--19},
  year={2021},
  publisher={Nature Publishing Group}
}

@article{rust2010selectivity,
  title={Selectivity and tolerance (“invariance”) both increase as visual information propagates from cortical area V4 to IT},
  author={Rust, Nicole C and DiCarlo, James J},
  journal={Journal of Neuroscience},
  volume={30},
  number={39},
  pages={12978--12995},
  year={2010},
  publisher={Soc Neuroscience}
}

@article{vermaercke2014functional,
  title={Functional specialization in rat occipital and temporal visual cortex},
  author={Vermaercke, Ben and Gerich, Florian J and Ytebrouck, Ellen and Arckens, Lutgarde and Op de Beeck, Hans P and Van den Bergh, Gert},
  journal={Journal of neurophysiology},
  volume={112},
  number={8},
  pages={1963--1983},
  year={2014},
  publisher={American Physiological Society Bethesda, MD}
}

@article{vinken2016neural,
  title={Neural representations of natural and scrambled movies progressively change from rat striate to temporal cortex},
  author={Vinken, Kasper and Van den Bergh, Gert and Vermaercke, Ben and Op de Beeck, Hans P},
  journal={Cerebral Cortex},
  volume={26},
  number={7},
  pages={3310--3322},
  year={2016},
  publisher={Oxford University Press}
}

@article{walker2019inception,
  title={Inception loops discover what excites neurons most using deep predictive models},
  author={Walker, Edgar Y and Sinz, Fabian H and Cobos, Erick and Muhammad, Taliah and Froudarakis, Emmanouil and Fahey, Paul G and Ecker, Alexander S and Reimer, Jacob and Pitkow, Xaq and Tolias, Andreas S},
  journal={Nature neuroscience},
  volume={22},
  number={12},
  pages={2060--2065},
  year={2019},
  publisher={Nature Publishing Group}
}

@article{panzeri1996analytical,
  title={Analytical estimates of limited sampling biases in different information measures},
  author={Panzeri, Stefano and Treves, Alessandro},
  journal={Network: Computation in neural systems},
  volume={7},
  number={1},
  pages={87},
  year={1996},
  publisher={IOP Publishing}
}

@article{gonzalez2008digital,
  title={Digital image processing, prentice hall},
  author={Gonzalez, Rafael C and Woods, Richard E},
  journal={Upper Saddle River, NJ},
  year={2008}
}

@article{panzeri2007correcting,
  title={Correcting for the sampling bias problem in spike train information measures},
  author={Panzeri, Stefano and Senatore, Riccardo and Montemurro, Marcelo A and Petersen, Rasmus S},
  journal={Journal of neurophysiology},
  volume={98},
  number={3},
  pages={1064--1072},
  year={2007},
  publisher={American Physiological Society}
}

@article{brincat2018gradual,
  title={Gradual progression from sensory to task-related processing in cerebral cortex},
  author={Brincat, Scott L and Siegel, Markus and von Nicolai, Constantin and Miller, Earl K},
  journal={Proceedings of the National Academy of Sciences},
  volume={115},
  number={30},
  pages={E7202--E7211},
  year={2018},
  publisher={National Acad Sciences}
}

@article{durand2016comparison,
  title={A comparison of visual response properties in the lateral geniculate nucleus and primary visual cortex of awake and anesthetized mice},
  author={Durand, S{\'e}verine and Iyer, Ramakrishnan and Mizuseki, Kenji and de Vries, Saskia and Mihalas, Stefan and Reid, R Clay},
  journal={Journal of Neuroscience},
  volume={36},
  number={48},
  pages={12144--12156},
  year={2016},
  publisher={Soc Neuroscience}
}

\end{document}